\documentclass[notitlepage, twocolumn, prl, aps, 10pt, superscriptaddress]{revtex4-1}
\usepackage{graphicx}
\usepackage{epstopdf}
\usepackage{tabularx}
\usepackage{easymath}

\begin{document}
\title{Electrokinetic Control of Viscous Fingering}
\author{Mohammad Mirzadeh}
\affiliation{Department of Chemical Engineering, Massachusetts Institute of Technology, MA 02139.}
\author{Martin Z. Bazant}
\email[]{bazant@mit.edu}
\affiliation{Department of Chemical Engineering, Massachusetts Institute of Technology, MA 02139.}
\affiliation{Department of Mathematics, Massachusetts Institute of Technology, MA 02139.}
\date{\today}

\begin{abstract} We present a theory of the interfacial stability of two immiscible electrolytes
under the coupled action of pressure gradients and electric fields in a Hele-Shaw cell or porous
medium. Mathematically, our theory describes a phenomenon of ``Vector Laplacian Growth'', in which
the interface moves in response to the gradient of a vector-valued potential function through a
generalized mobility tensor. Physically, we extend classical Saffman-Taylor problem to electrolytes
by incorporating electrokinetic phenomena. A surprising prediction is that viscous fingering can be
controlled by varying the injection ratio of electric current to flow rate. Beyond a critical
injection ratio, stability depends only upon the relative direction of flow and current, regardless
of the viscosity ratio. Possible applications include porous materials processing, electrically
enhanced oil recovery, and electrokinetic remediation of contaminated soils. \end{abstract}

\maketitle

Interfacial instability is the precursor to pattern formation in a variety of physical and chemical
processes \cite{Kessler1988Pattern-selection-in,pelce2012dynamics}. This fascinating topic covers a
broad range of phenomena such as dendritic growth due to the Mullins-Sekerka instability in
solidification \cite{mullins1963morphological,*Mullins1964,langer1980instabilities}, fractal growth
due to diffusion-limited aggregation \cite{Witten1983Diffusion-limited-ag} or metal
electrodeposition \cite{brady1984fractal} in fluid flows~\cite{Bazant2003}, crease formation and
wrinkling of combustion fronts due to the Darrieus-Landau instability
\cite{darrieus1938propagation,*landau1944theory,sivashinsky1983instabilities,*matalon2007intrinsic},
and viscous fingering in Hele-Shaw cells~\cite{Bensimon1986Viscous-flows-in-two} and porous
media~\cite{Homsy1987Viscous-fingering-in} due to the Saffman-Taylor instability
\cite{Saffman1958The-penetration-of-a,Chuoke1959The-instability-of-s}.

Interfacial instabilities are usually undesirable, but difficult to control. In secondary oil
recovery, viscous fingering of injected liquids leads to nonuniform displacement and residual
trapping of oil~\cite{Homsy1987Viscous-fingering-in,gorell1983theory}, and dendritic growth is a
major safety concern for metal anodes in rechargeable batteries \cite{xu2014lithium}. There are
signs, however, that instability may be avoided if the interface is driven by multiple opposing
forces. For instance, it was recently observed that dendritic growth can be suppressed in charged
porous media \cite{han2016dendrite} if preceded by deionization shock wave
\cite{mani2011deionization}, whose stable propagation in cross flow also enables water purification
by shock electrodialysis~\cite{deng2013overlimiting,schlumpberger2015scalable}.

Here, we consider the interfacial stability of two immiscible electrolytes in a Hele-Shaw cell
where the interface is set into motion by both the pressure-driven and electro-osmotic flows.
Remarkably, we find that electrokinetic coupling influence interfacial stability and, under certain
conditions, can eliminate viscous fingering. This phenomenon illustrates the rich physics of
``Vector Laplacian Growth'' (VLG), a general mathematical model of interfacial dynamics driven by
the gradient of a vector-valued potential function through a generalized mobility tensor. The
``one-sided'' VLG model (with field gradients only on one side of the interface) is known to be
unstable, leading to fractal patterns, during growth \cite{Bazant2003} and stable, resulting in
smooth collapse, during retreat \cite{Bazant2006a,Rycroft2015}. Our theory shows that stable growth
is also possible, if field gradients exist on both sides of the interface.
\begin{figure}
  \centering
  \includegraphics[width = \columnwidth]{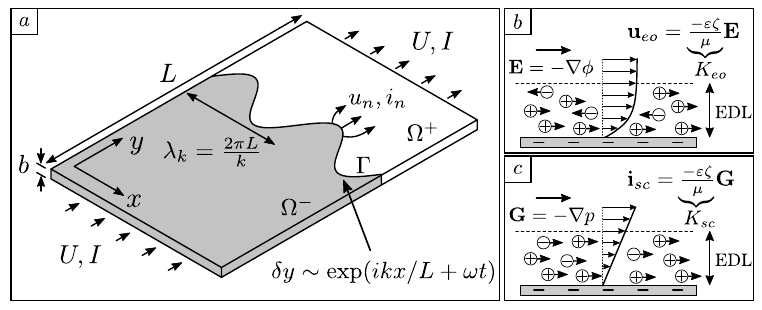}
  \caption{Schematic of flow in a rectangular Hele-Shaw cell. (a) The interface ($\Gamma$) between
  two immiscible electrolytes moves under the coupled action of pressure gradient and electric field
  as described by the electrokinetic response of the cell. (b) In addition to the pressure-driven
  flow, the electric field exerts a net force on ions in the Electric Double Layer (EDL), resulting
  in the electro-osmotic flow. (c) Similarly, in addition to the Ohmic current driven by the
  electric field, the pressure-driven flow advects charges in the EDL, resulting in the streaming
  current.}
  \label{fig:schematic}
\end{figure}

In the classical viscous fingering problem, the fluid flow in a Hele-Shaw cell can be approximated
as quasi two-dimensional if the cell gap, $b$, is much smaller than the lateral dimension, $L$ (see
fig \ref{fig:schematic}). In this case, the gap-averaged velocity of each fluid is given by:
\be
\mathbf{u}^\pm = -\frac{b^2}{12 \mu^\pm} \nabla p^\pm, \hspace{5 mm} \nabla \cdot \mathbf{u}^\pm = 0 \label{eq:velocity}
\ee
where `$-$' and `$+$' superscripts denote invading and receding fluids, $\mu$ and $p$ are viscosity and pressure of each fluid, and $\nabla$ is the in-plane gradient operator. At the interface, the pressure jump is given by the Young-Laplace equation, while the normal velocity is continuous:
\be
\jump{p} = \gamma \kappa, \hspace{5 mm} \jump{\normal{\mathbf{u}}} = 0, \label{eq:jump}
\ee 
where $\jump{a} \equiv a^+ - a^-$ denotes the jump of variable `$a$' across the interface,
$\gamma$ is the surface tension, and $\kappa$ is the in-plane curvature. More generally, these
conditions must be modified to take the finite lubrication film thickness into account if the
receding fluid is perfectly wetting \cite{park1984two,Homsy1987Viscous-fingering-in}. The interface
moves with the local fluid velocity, 
\be
\dx{\mathbf{x}}{t} =  (\normal{\mathbf{u}}) \nhat,
\label{eq:kinematic}
\ee
and the far-field flow is uniform.

Linear stability analysis of Eqs.\ \eqref{eq:velocity}--\eqref{eq:kinematic}, initiated by Chuoke
\etal \cite{Chuoke1959The-instability-of-s} and Saffman and Taylor
\cite{Saffman1958The-penetration-of-a}, reveals that stable displacement is only possible if the
advancing fluid is more viscous:
\be
\mbox{ Stable: } \ \ M = \frac{\mu^-}{\mu^+} > 1.
\label{eq:def M}
\ee 
In the opposite case, $M<1$, the  interface is unstable to perturbations of sufficiently long
wavelength, and the less viscous fluid forms ``fingers" of lower resistance through the more viscous
fluid. Specifically, the growth rate, $\omega$, of a normal mode $\delta y \sim \exp(ikx/L+\omega
t)$ satisfies the dispersion relation
\cite{Homsy1987Viscous-fingering-in,Bensimon1986Viscous-flows-in-two}:
\be
\omega = \frac{k}{L}\left(U\frac{\mu^+-\mu^-}{\mu^++\mu^-} - \frac{\gamma b^2 k^2}{12 L^2 \left(\mu^++\mu^-\right)}\right),
\label{eq:omega_vis}
\ee
where $k$ is the wavenumber. Perturbation wavelengths longer than $\lambda_{cr} = \pi b \sqrt{\gamma/3 U (\mu^+-\mu^-)}$ are unstable, and the maximum growth rate arises for $\lambda_m = \sqrt{3} \lambda_{cr}$. 

Most materials naturally acquire charge in aqueous solutions from the dissociation of surface
groups, such as silanol \cite{kirby2004zeta,hunter2013zeta}, for glass in Hele-Shaw cells or
silicate minerals in underground reservoirs. The screening of surface charge by mobile ions leads to
the formation of electric double layers (EDL) and associated electrokinetic phenomena
\cite{russel1989colloidal,*lyklema1995fundamentals}. An electric field parallel to the charged
surface acts on EDL charge to drive ``electro-osmotic'' flow $\mathbf{u}_{eo}$, while
pressure-driven flow drives ``streaming current" $\mathbf{i}_{sc}$ due to the advection of EDL
charge (Fig.\ \ref{fig:schematic}). For typical situations of fixed surface charge, the
electrokinetic response is linear in the driving forces, \ie $\mathbf{u}_{eo} = -K_{eo} \nabla \phi$
and $\mathbf{i}_{sc} = -K_{sc} \nabla p$, where $\phi$ is the electrostatic potential. The
electro-osmotic mobility, $K_{eo}$, and the streaming conductance, $K_{sc}$, satisfy Onsager's
reciprocal relation \cite{onsager1931reciprocal-I,Bazant2016}, $K_{eo} = K_{sc}$, and for thin EDL
(gaps, $b \sim 0.1-1 \: \text{mm}$, much larger than the EDL thickness, $\lambda_D \sim 1-10 \:
\text{nm}$), are given by the Helmholtz-Smoluchowski relation \cite{lyklema1995fundamentals},
$K_{eo} = -\varepsilon \zeta/\mu$, where $\varepsilon$ is the electrolyte permittivity and $\zeta$
is the potential difference across the EDL \cite{lyklema1995fundamentals, russel1989colloidal,
hunter2013zeta}.

When linear electrokinetic phenomena are considered, a VLG model can thus be written in terms of a
tensorial flux, $\mathbf{F} = (\mathbf{u}, \mathbf{i})^\mathsf{T}$, proportional to the gradient of
a vector-valued potential, $\mathbf{\Phi} = (p, \phi)^\mathsf{T}$:
\be
  \mathbf{F}^\pm = -\mathbb{K}^\pm \nabla \mathbf{\Phi}^\pm, \hspace{5 mm} \nabla \cdot \mathbf{F}^\pm = \mathbf{0},
  \label{eq:vectorform}
\ee
where $\mathbb{K}$ is the electrokinetic mobility tensor:
\be
\mathbb{K} =
\left(\begin{array}{cc}
      K_h    & K_{eo} \\
      K_{eo} & K_e
      \end{array}
\right).
\label{eq:mobility tensor}
\ee 
$K_h = b^2/12 \mu$ is the hydraulic Darcy conductivity, and $K_e = \sigma$ is the electrical
Ohmic conductivity of the cell. The Second Law of Thermodynamics requires positive definite
$\mathbb{K}$ to ensure positive dissipation rate \cite{de2013non,Peters2016}, i.e.:
\be
-\nabla \mathbf{\Phi} \cdot \mathbf{F} = \nabla \mathbf{\Phi}^\mathsf{T} \mathbb{K} \nabla \mathbf{\Phi} > 0.
\label{eq:spd}
\ee 
At the interface, the pressure and total velocity satisfy the jump conditions given by Eq.\
\eqref{eq:jump}, while the potential and normal component of the total current are continuous, which
can be compactly expressed  as:
\be
\jump{\mathbf{\Phi}} = 
(\gamma \kappa , 0)^\mathsf{T}
, \hspace{5 mm} 
\jump{\normal{\mathbf{F}}} = \mathbf{0}.
\label{eq:total jump}
\ee 
Far from the interface in a planar geometry, the fluxes are assumed to be uniform,
$\lim_{y\rightarrow\pm\infty} \mathbf{F}_y = \mathbf{F}_\infty = (U, I)^\mathsf{T}$. Equations
\eqref{eq:vectorform} and \eqref{eq:total jump}, along with the kinematic condition
\eqref{eq:kinematic}, determine the interface motion.

As for classical problem, we consider the linear stability of a planar interface subjected to a
sinusoidal perturbation, $\delta y \sim \exp(ikx/L+\omega t)$, and seek solutions of the form
$\mathbf{\Phi}^\pm = \mathbf{\Phi}^\pm_0 + \epsilon \mathbf{\Phi}^\pm_1$ in the limit of $\epsilon
\ll 1$. From Eq.\ \eqref{eq:vectorform}, the base state is linear, \ie $\mathbf{\Phi}^\pm_0 =
-\mathbb{K}^{\pm^{-1}} \mathbf{F}_\infty (y - Ut)$, while $\mathbf{\Phi}^\pm_1 =
\mathbf{A}^\pm_1\exp(ikx/L+\omega t)\exp(\mp k(y-Ut)/L)$, where $\mathbf{A}^\pm_1$ are evaluated
using the jump condition \eqref{eq:total jump}. Applying the kinematic condition
\eqref{eq:kinematic} then yields the growth rate:
\be
\omega = \frac{k}{L}\left(F - \gamma G \frac{k^2}{L^2}\right),
\label{eq:omega_general}
\ee
where $F$ and $G$ are given by
\begin{gather}
\nonumber F = U\frac{\jump{K_{eo}}\{K_{eo}\}-\jump{K_h}\{K_e\}}{\det \{\mathbb{K}\}}+2I\frac{K_h^+K_{eo}^--K_h^-K_{eo}^+}{\det \{\mathbb{K}\}}, \label{eq:F}\\
G = \frac{K_h^+ \det \mathbb{K}^- + K_h^- \det \mathbb{K}^+}{\det \{\mathbb{K}\}}, \label{eq:G}
\end{gather}
and $\{a\} \equiv a^+ + a^-$. Note that the classical dispersion relation \eqref{eq:omega_vis} is
recovered in the absence of electrokinetic phenomena, $K_{eo}^\pm = 0$. From the Second Law
\eqref{eq:spd}, it follows that $G>0$, ensuring that surface tension effects are stabilizing.
Therefore, $F<0$ is a \emph{sufficient} condition for stability. For $F>0$, a perturbation of
wavelength longer than $\lambda_{cr} = 2\pi\sqrt{\gamma G/F}$ is unstable, and $\lambda_m = \sqrt{3}
\lambda_{cr}$ is the most unstable wavelength. 

\begin{figure}
\centering
\includegraphics[width = \columnwidth]{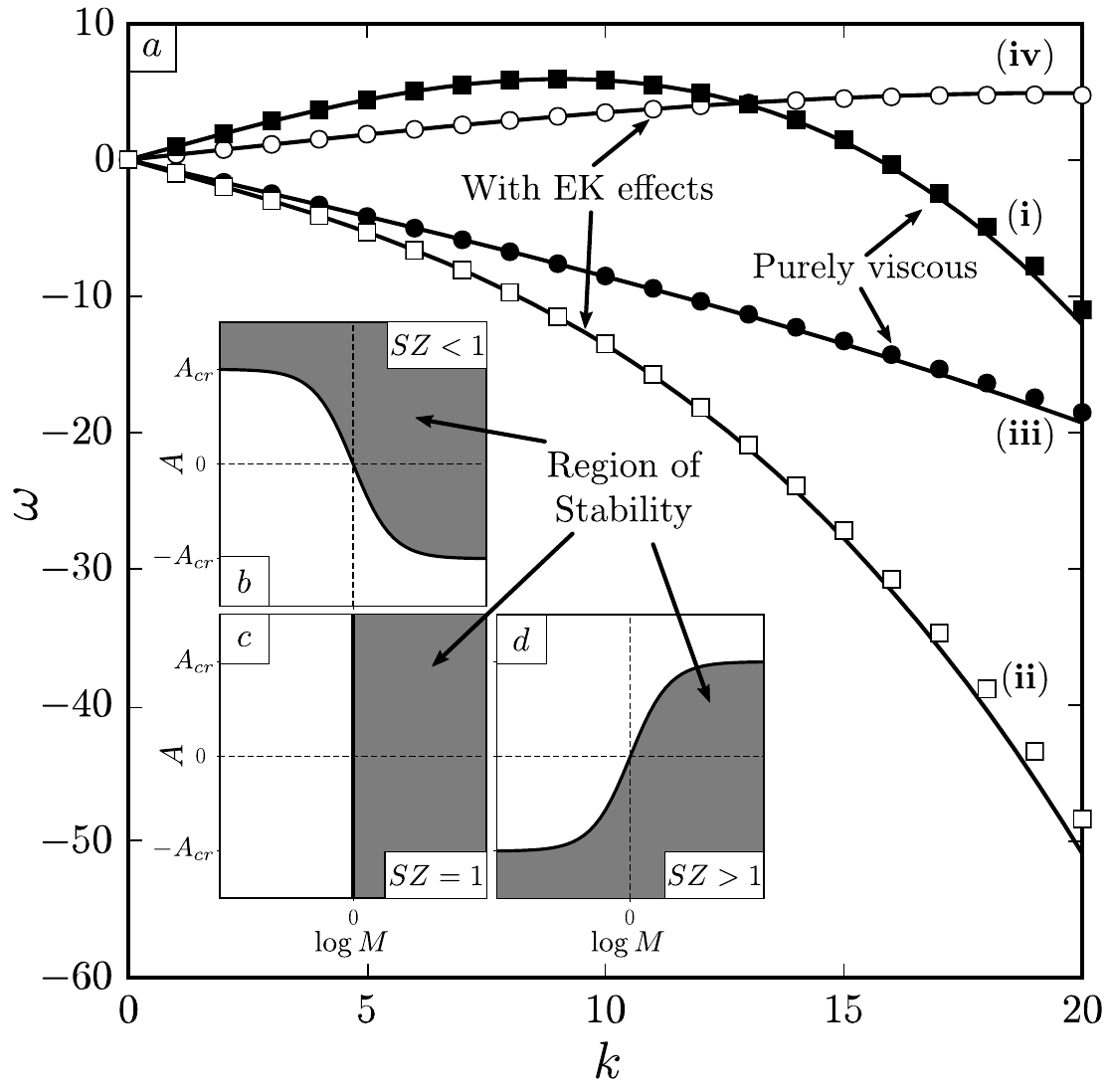}
\caption{Linear stability analysis of a planar interface. (a) The non-dimensional growth rate
$\omega$ (scaled with $U/L$) versus the wave number $k$. Solid lines represents the theory (see eq.\
\eqref{eq:omega_general}) while symbols are numerically computed growth rates, obtained by evolving
a small-amplitude initial perturbation ($\epsilon = 10^{-3}$) for each wavenumber. Shown are
classical results (no EK effects) for (\textbf{i}) unfavorable ($M = 0.01$) and (\textbf{iii})
favorable ($M=10$) viscosity ratios. When electrokinetic effects are present, stability can be
manipulated by adjusting the injection ratio, resulting in either (\textbf{ii}) suppression of
viscous fingering ($M = 0.01, SZ = R = 100, A \approx -1.98$), or (\textbf{iv}) electrokinetic
fingering ($M = SZ = R = 10, A \approx 1.45$). Here, $R=\sigma^-/\sigma^+$ is the conductivity ratio
and the remaining parameters are defined via equation \eqref{eq:ratios}. In all cases, the effective
Capillary number, $\text{Ca} = 12 L^2 U \mu^+/\gamma b^2$, is set to $250$. (b--d) The shaded area
illustrates the region of stability as approximated by equation \eqref{eq:stability_nondim}.
Interestingly, this region is symmetrical around $SZ=1$, for which classical results are recovered.
}
\label{fig:instability}
\end{figure}

\begin{figure}
\centering
\includegraphics[width = \columnwidth]{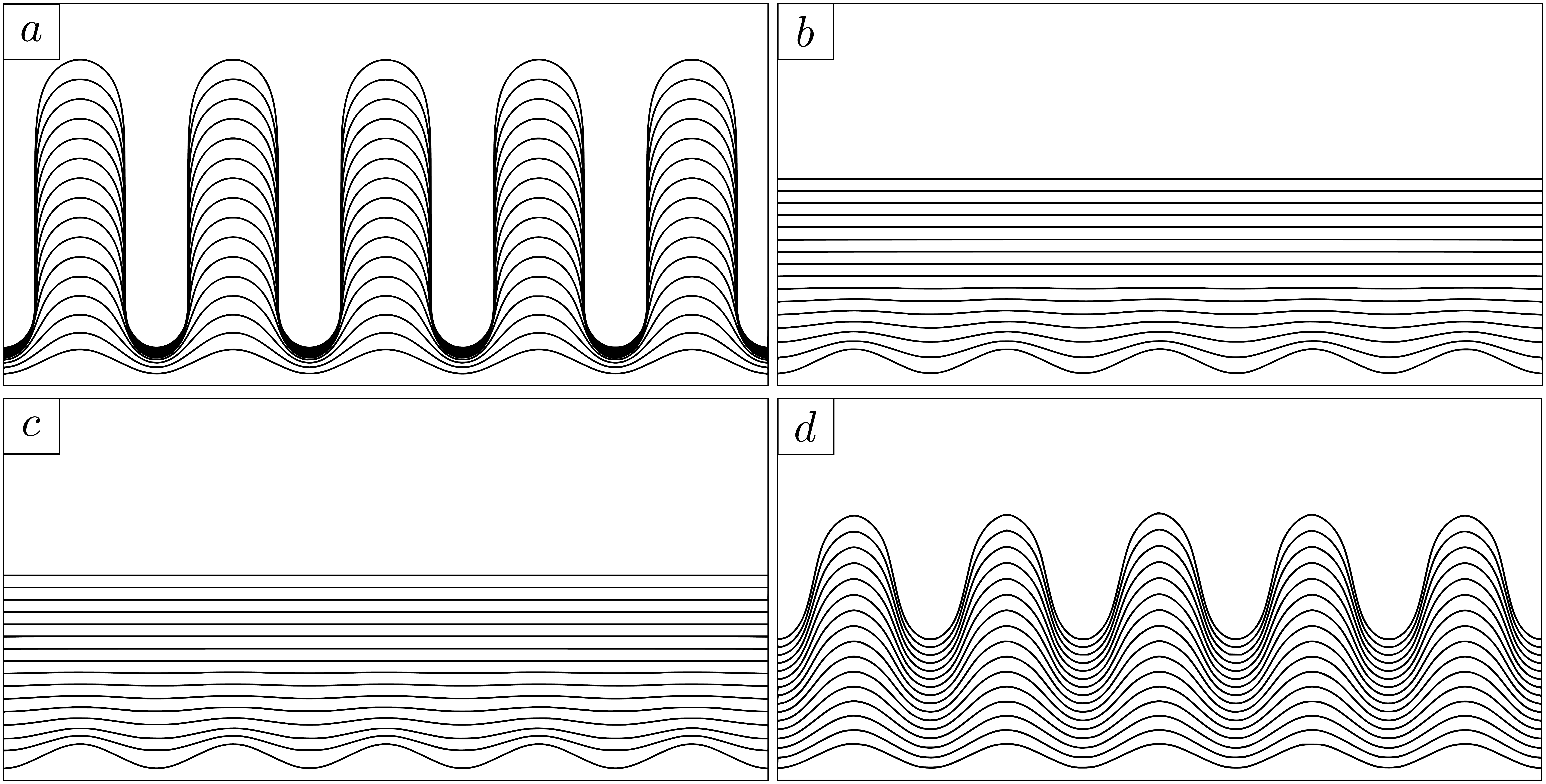}
\caption{Numerical simulation of the interfacial dynamics for an initial perturbation. The interface
moves upward and its location is drawn at equal time intervals. (a) Viscous fingering for an
unfavorable viscosity ratio $M = 0.01$ and (b) its suppression using negative injection ratio (same
parameters as in fig.\ 2a-\textbf{ii}). (c) Stable displacement for favorable viscosity ratio $M =
10$ and (d) formation of electrokinetic fingering with positive injection ratio (same parameters as
in fig.\ 2a-\textbf{iv}). }
\label{fig:simulation planar}
\end{figure} 

To simplify equation \eqref{eq:G}, we note that the electrokinetic coupling coefficient
\cite{VanDerHeyden2006,*VanDerHeyden2007}, $\alpha = K_{eo}^2/K_h K_e$, is typically small, while $0
\le \alpha < 1$ from Eq.\ \eqref{eq:spd}. For $\alpha \ll 1$, the critical wavelength may be
approximated as:
\be
\lambda_{cr} = \pi b \sqrt{\frac{\gamma}{3U\jump{\mu} + 6I \jump{\varepsilon\zeta}/\{\sigma\}}}.
\label{eq:lambda_cr}
\ee 
While the classical instability is controlled by the viscosity ratio (Eq.\ \eqref{eq:def M}),
our theory predicts that the injection ratio, $I/U$, can be tuned \emph{independently} to control
interfacial  stability (Fig.\ \ref{fig:instability}a):
\be
\mbox{Stable:} \ \ \frac{M-1}{M+1} > A \frac{SZ-1}{SZ+1},
\label{eq:stability_nondim}
\ee
in terms of the following dimensionless ratios:
\be
S = \frac{\varepsilon^-}{\varepsilon^+}, \hspace{3 mm}
Z = \frac{\zeta^-}{\zeta^+}, \hspace{3 mm}
M = \frac{\mu^-}{\mu^+}, \hspace{3 mm}
A = \frac{I(-\overline{\varepsilon\zeta})}{U \bar{\mu} \bar{\sigma}}, 
\label{eq:ratios}
\ee 
where the over-bar indicates average values, \eg $\overline{\varepsilon\zeta} =
\left(\varepsilon^+\zeta^+ + \varepsilon^-\zeta^-\right)/2$. Electrokinetic effects require $SZ \ne
1$, and stability is possible if the injection ratio is larger than
$\left|(M-1)(SZ+1)/(SZ-1)(M+1)\right|$, and has the ``correct'' sign, depending on the magnitude of
$SZ$ (see Fig.\ \ref{fig:instability}). Above a critical injection ratio, $A_{cr} =
\left|(SZ+1)/(SZ-1)\right|$, stability is entirely determined by the sign of injection ratio and is,
remarkably, independent of the viscosity ratio. Physically, negative injection ratios denote
opposite direction of current and flow. These observations are illustrated in figure
\ref{fig:instability}b--d.

Motivated by secondary oil recovery, it is interesting to consider the limit when $M \ll 1$, $SZ \gg
1$, and $R = \sigma^- / \sigma^+ \gg 1$, \eg when water is pushing oil toward extraction wells. In
this case, negative current injection shifts the critical wavelength to longer values and reduces
viscous fingering. Stable displacement is possible if $I>I_{cr}$, where:
\be
I_{cr} \approx \frac{U}{2}\frac{\mu^o}{\mu^w} \frac{\sigma^w}{K_{eo}^w}.
\label{eq:I_cr}
\ee 
For $U = 1 \:\text{mm}/\text{min}$ in a $1 \:\text{mM}$ KCl solution with $\zeta = -50 \:
\text{mV}$ and $\mu^w/\mu^o = 0.1$, the  critical current is fairly small, $I_{cr} \sim 4 \:
\text{mA}/\text{cm}^2$, but a large critical electric field is required to  drive this current
across the poorly conducting oil region: 
\be
E_{cr} \approx \frac{U}{4}\frac{\mu^o}{\mu^w}\frac{\sigma^w}{\sigma^o} \frac{1}{K_{eo}^w}.
\ee 
Even for a modest value of $\sigma^w/\sigma^o = 10$, complete suppression of viscous fingering
requires $E_{cr} \sim 150 \: \text{V}/\text{cm}$. The required voltage could be lowered by reducing
the conductivity ratio or electrode separations. Nonetheless, partial stabilization (with enhanced
oil recovery) is still viable with electric fields below the critical value.

To further support our theory, we numerically solve the VLG model using the Voronoi Interface Method
\cite{Guittet2015} to discretize the conservation equations \eqref{eq:vectorform} subjected to the
interface jump conditions \eqref{eq:total jump} while utilizing the level-set framework
\cite{Osher1988,*osher2006level,*sethian1999level} to represent the moving interface. Furthermore,
we use dynamically adaptive quadtree grids \cite{MG07A} as well as parallel algorithms
\cite{Mirzadeh2016} for fast and high-fidelity simulations. Figure \ref{fig:simulation planar}
illustrates the interfacial dynamics of an initial perturbation for unfavorable, \ref{fig:simulation
planar}-(a,b), and favorable, \ref{fig:simulation planar}-(c,d), viscosity ratios. As predicted,
interfacial stability can be manipulated by adjusting the injection ratio.

\begin{figure}
\centering
\includegraphics[width = \columnwidth]{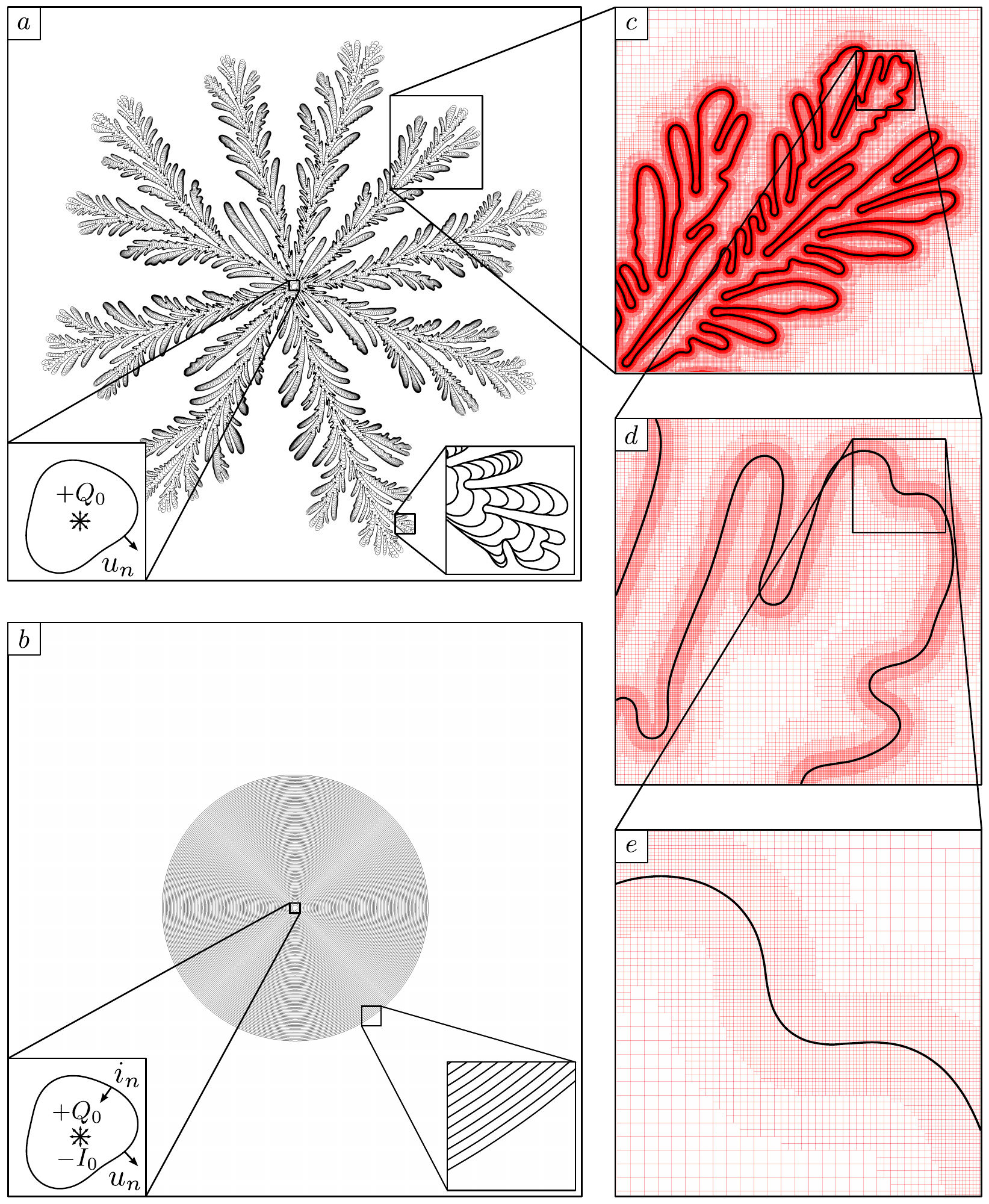}
\caption{Numerical simulation of the interfacial dynamics in a radial Hele-Shaw cell. (a) An initial
``bubble'' (left inset) separates two immiscible ($M = 0.01$) and grows outward due to a positive
flow source in the middle. The interface location, drawn at equal time intervals (right inset),
reveals a complex pattern due to successive growth and tip-splitting. (b) The viscous fingering is
entirely suppressed by injecting electric current in the opposite direction, resulting in a uniform
circular growth (same parameters as in fig.\ 2a-\textbf{ii}). (c-e) Snapshot of the final pattern in
(a), illustrating the level of details that is captured in the simulation.}
\label{fig:simulation radial}
\end{figure}

It is straight-forward to extend the analysis for the radial Hele-Shaw cell geometry
\cite{Paterson1981Radial-fingering-in-,Homsy1987Viscous-fingering-in} where an invading fluid is
injected at a point to push the second fluid outward. If the interface is initially assumed to be
circular, the growth rate of an azimuthal perturbation of the form $\delta r \sim
\exp\left(ik\theta+\omega t\right)$, is given via:
\be
\omega = -\frac{U}{r} + \frac{k}{r}\left(F - \gamma G \frac{(k^2 - 1)}{r^2}\right),
\ee 
where $r$ is the initial radius and $F$ and $G$ are still given by equation \eqref{eq:G}. Once
again, $F<0$ is a sufficient condition for stability. Therefore, the stability estimate in equation
\eqref{eq:stability_nondim} could be used in radial geometry if velocity, $U$, and current density,
$I$, are replaced by the total flow rate, $Q_0$, and total current, $I_0$, respectively. Figure
\ref{fig:simulation radial} illustrates numerical simulation of interface evolution in a radial
Hele-Shaw cell geometry for an unstable viscosity ratio of $M=0.01$. The instability is entirely
suppressed when current is injected in the opposite direction.

The possibility of manipulating interfacial instabilities is quite exciting. The idea of
controlling viscous fingering using cell geometry has been recently discussed
\cite{Al-Housseiny2012Control-of-interfaci,*Al-Housseiny2013a}, and our framework introduces many
other degrees of freedom, such as the placement of electrodes, dielectric or conducting boundaries,
and surface coatings or gate voltages to modify local zeta potentials. For a given geometry,
dynamical control of fingering instabilities may also be possible, by adjusting potentials and
pressures with real-time feedback from currents and flow rates.

Hele-Shaw cell experiments could be used to check these predictions and test the validity of our
assumptions. Since electrokinetic phenomena depend on which liquid is in contact with the surface,
it may be necessary to extend the model for lubrication films and gravity
currents~\cite{eames2005effect}, which would require more complicated depth-averaging and
electrokinetics at the liquid-liquid interface \cite{Pascall2017}, perhaps amenable to
conformal-map dynamics~\cite{Bazant2003,robinson2017conformally}. We have also neglected non-linear
electro-hydrodynamic effects \cite{Melcher1969,*Melcher1981,*melcher1963field} which might cause
interfacial instabilities at higher electric fields in large channels
\cite{TaylorAN2017,Lin2004,*Chen2005,Druzgalski2013a}. We also assume finite electrical resistance
in each phase, which could exclude traditional liquid pairs, such as water/silicon oil and
air/glycerol, although some poorly conducting regions may have sufficient ionic or electronic
conductivity to pass at least a transient current consistent with the model. The model could be
extended to include interfacial capacitance, and the resulting ``RC time" for charge accumulation
might be longer than the instability growth time, especially for large resistive domains.

Our model is directly applicable to interfaces between two immiscible electrolyte solutions (ITIES)
that support charge-transfer reactions~\cite{girault1987electrochemistry}. Examples include aqueous
electrolytes, \eg LiCl, in contact with solutions of lipophilic salts in organic solvents, \eg
TBATPB in nitrobenzene \cite{Senda1991, *girault1993charge, *Samec2004,
*girault2010electrochemistry}. Recent interest in ITIES was spurred by
electro-wetting~\cite{monroe2006electrowetting} for electro-variable
optics~\cite{edel2016fundamentals}, but tunable fingering under confinement could lead to different
applications.

Although our theory is for immiscible electrolytes, it may also describe diffuse interfaces
involving strong ion concentration gradients, \eg deionization shocks in charged porous media
\cite{mani2011deionization}or pH fronts in electrokinetic remediation of contaminated soil
\cite{Probstein2011,*Shapiro1993}. Since the $\zeta$-potential is a function of pH and salt
concentration \cite{kirby2004zeta, hunter2013zeta}, it may also be possible to observe some of the
stabilizing effects with miscible solutions, perhaps in charged porous media such as glass frit or
Hele-Shaw cells packed with silica beads. Finally, we caution that viscous fingering is more
complicated in porous media than in Hele-Shaw cells, due to permeability variations, capillary
effects, and surface wettability \cite{Lenormand1988a,Zhao2016}. Since electrokinetic couplings
derive from surfaces, we expect strong dependence on surface wettability whereas permeability
variations might have limited impact due to disproportionate scaling of hydraulic and
electro-osmotic mobilities with the pore size, possibly resulting in a more uniform displacement.
Nonetheless, further investigation is required to quantify the degree to which electrokinetic
phenomena can control interfacial stability in porous media.

This work was supported by a seed grant from the MIT Energy Initiative with computational resources
from the Texas Advanced Computing Center (TACC) at The University of Texas at Austin and the
Extreme Science and Engineering Discovery Environment (XSEDE), supported by grant ACI-1548562 from
the National Science Foundation. The authors thank Amir A.\ Pahlavan and Charles W.\ Monroe for
useful discussions.

\bibliographystyle{apsrev4-1}
\bibliography{refs}
\end{document}